\documentclass[11pt,a4paper,final]{article}
\usepackage{amsmath}
\usepackage{amsfonts}
\usepackage{amssymb}
\usepackage{graphicx}
\usepackage{algorithm} 
\usepackage{algpseudocode}
\usepackage{natbib}
\usepackage[margin=1.in]{geometry}
\usepackage{color}
\usepackage{bm}
\usepackage{hyperref}
\usepackage{amsthm}
\usepackage{authblk}
\usepackage{ulem}

\begin{document}

\title{Rejoinder for ``Probabilistic Integration:\\ A Role in Statistical Computation?''}

\author{ Fran\c{c}ois-Xavier Briol (Imperial College London), Chris J. Oates (Newcastle University), Mark Girolami (Imperial College London), Michael A. Osborne (University of Oxford) \& Dino Sejdinovic (University of Oxford).}

\maketitle

\begin{abstract}
This article is the rejoinder for the paper ``Probabilistic Integration: A Role in Statistical Computation?'' to appear in Statistical Science with discussion \citep{Briol2015PI}. We would first like to thank the reviewers and many of our colleagues who helped shape this paper, the editor for selecting our paper for discussion, and of course all of the discussants for their thoughtful, insightful and constructive comments. In this rejoinder, we respond to some of the points raised by the discussants and comment further on the fundamental questions underlying the paper: 
\begin{center}
$\bullet$ Should Bayesian ideas be used in numerical analysis? \\ $\bullet$ If so, what role should such approaches have in statistical computation?
\end{center}
\end{abstract}

\section*{Bayesian Numerical Methods}

Numerical analysis is concerned with the approximation of typically high or infinite-dimensional mathematical quantities using discretisations of the space on which these are defined. Examples include integrands, and as a consequence their corresponding integrals, or the solutions of differential equations. Different discretisation schemes lead to different numerical algorithms, whose stability and convergence properties need to be carefully assessed. Such numerical methods have undeniably played, and continue to play, an important role in the implementation of statistical methods, from the computation of intractable integrals in Bayesian statistics to the computation of estimators using optimisation routines in frequentist statistics. 

In essence, a Bayesian numerical method models the quantity of interest as a latent variable. A prior model is posited on this quantity, then the discretisation of the space is used as data to turn the prior distribution into a posterior distribution. The prior models used are often constructed over functions spaces, and the literature on Bayesian numerical methods therefore draws much inspiration from the work on Bayesian nonparametrics, such as Gaussian processes.

Since their inception in the seventies and eighties, Bayesian numerical methods have initiated much debate as to the foundations of computation and approximation of unknown functions. In this regard, Bayesian numerical methods have initiated fresh considerations of what it means to ``know a function'' \citep{Diaconis1988}. Several important questions have been raised - many of which are highlighted by the discussants of this paper. Are Bayesians simply reinventing the bread and butter of numerical analysis using a different scientific language? How should a prior model on such an abstract quantity be selected? What interpretation should be given to the posterior?

These types of questions have a different emphasis depending on the priorities of the varying scientific communities that are engaged in this debate.  We hope that the present paper and contributed comments can bring some focus to this discussion, and help clarify the advantages and disadvantages of the approach. The answer from the discussants to our first central question can be summarised as ``Yes, but...", which we believe is representative of the overall thoughts of the various research communities represented in this field. There are of course many caveats with these methods, and they are not reasonably expected to supplant numerical analysis, but we will argue that some of their advantages can be significant for certain application areas.

\section*{The Quantification of Uncertainty in Numerical Problems}

One motivation for the use of Bayesian methods in numerical analysis is the possibility of quantifying uncertainty emanating from discretisation error. Indeed, the posterior provides an entire distribution over the solution of the numerical problem which reflects any uncertainty remaining due to the fact that we have access to a finite amount of data, or in other words, that our observations of the quantity of interest are limited to the discretisation of the space. This is particularly appealing since this distribution can be used for several tasks, including to decide whether to refine the discretisation, and if this is of interest, to construct an experimental-design strategy to do so.  

However, there are several technical challenges remaining to guarantee that our quantification of uncertainty is valid. First, we must justify our choice of prior model using knowledge of the function derived from domain expertise. This can clearly be a difficult task. Second, we also need to verify the consistency of the method, a problem which, unlike in finite-dimensional spaces, can be technically challenging \citep{Diaconis1986,Owhadi2015Brittleness}. Finally, even though the confidence intervals are valid from a Bayesian viewpoint (in the sense that they reflect our remaining uncertainty given our prior and the data), we may want to verify that the posterior is calibrated from a frequentist viewpoint; once again a non-trivial matter \citep{Cox1993}.

In the paper, we illustrated these issues for an algorithm called Bayesian quadrature/cubature, which tackles numerical integration by modelling the integrand using a prior (usually a Gaussian process), and using $n$ function evaluations to obtain a posterior on this integrand, which itself induces a posterior on the value of the integral. Our paper studied the consistency problem for this algorithm, and provided asymptotic rates. It also studied the calibration of the algorithm on some synthetic problems, where it was demonstrated that the posterior had good frequentist coverage for large $n$ (but no theory for this was provided). 

In his comment, Owen argues that Monte Carlo methods are `unreasonably effective' thanks to the confidence intervals provided by the central limit theorem (CLT), which can be estimated at a faster rate in $n$ than the integral of interest. However, we point out that the CLT only provides confidence intervals which are valid in an asymptotic regime where $n \rightarrow \infty$. On the other hand, worst-case error used in the quasi-Monte Carlo (QMC) literature are also valid for finite $n$, but as pointed out by Owen, Hickernell and Jagadeeswaran, these are too pessimistic for most problems of interest. In comparison, contingent on a good methodology for eliciting a prior (a non-trivial task), the Bayesian approach can provide non-pessimistic uncertainty quantification for finite $n$. 

Recent developments in the theory of Bayesian nonparametrics also demonstrate that certain Bayesian models can be well-calibrated in a frequentist sense for large classes of target functions \cite{Szabo2015,Sniekers2015}. Interestingly, Stein and Hung pointed out further work on calibration in the computer experiments literature, where it was demonstrated that the data collection mechanism was of importance for good calibration \citep{Zhu2006}. Note that a common assumption in these is that of noisy observations, which is not common in numerical analysis where functions are usually evaluated without noise (although the noisy case may still be relevant for certain applications). Nevertheless, these papers give us hope that similar calibration results could also be derived for Bayesian numerical methods.

\section*{Bayesian Priors, Posteriors and the Difficulty of Working with Measures}

The main premise required to justify the use of Bayesian methods is that we are able to come up with a good prior model which represents all our information prior to observing any data. This is clearly somewhat unreasonable, and it is debatable to what extent it is possible to construct such priors for the particular task of solving numerical problems. This issue is highlighted by Owen's claim that credible intervals obtained from Bayesian quadrature are `not anybody's belief'. Similar criticisms are of course also valid for most Bayesian numerical methods, and it is clear that much research is needed to help in the construction of such priors.

In this paper, we studied the case of Gaussian process priors, which is by far the most common choice for Bayesian numerical methods. The main reason for this choice is that working with general measures defined on infinite-dimensional domains is a notably challenging problem and requires the use of advanced sampling methods \citep{Cotter2013}. On the other hand, working with Gaussian processes is much simpler since linear operations preserve Gaussianity, and so boil down to linear algebra. As pointed out by Owen, Hickernell and Jagadeeswaran, this may still usually require $O(n^3)$ operations, but many approximation schemes for Gaussian processes exist. Some cheaper exact schemes based on particular choices of point sets or covariance functions have also recently been developed for the specific case of Bayesian numerical methods \cite{Karvonen2017symmetric,Jagadeeswaran2018}. Furthermore, we believe that the approach based on Fourier transforms suggested by Stein and Hung could also be fruitful.

Gaussian priors also lead to Bayesian numerical methods which are closely related to traditional methods. Indeed, connections between the posterior variance of Bayesian methods using Gaussian measures and worst-case results in reproducing kernel Hilbert spaces (RKHS) are well known and studied in the information-based complexity literature; see \cite{Ritter2000}. One issue which often leads to much confusion when making use of this connection, and which is highlighted by Stein, Hung, Hickernell and Jagadeeswaran, is the fact that draws from the Gaussian process are not in the space reproduced by the covariance function of the Gaussian process. The connection between Gaussian processes and RKHSs is nonetheless deep and offers opportunities to transfer results between the two frameworks, as highlighted by \cite{Kanagawa2018}. This may be extremely relevant to problems in numerical analysis as demonstrated by the Brownian bridge example of Stein and Hung, and by the recent work of \citep{Kanagawa2016} on misspecified Gaussian process priors. 

For this reason, a central question is the following: How should we pick a prior for such an abstract quantity as an integrand? Clearly certain choices can be made based on prior knowledge of properties of the function, such as smoothness or periodicity, but we often still need to set certain amplitude or lengthscale parameters. These can be set using objective priors, as demonstrated in the paper and extended by Hickernell and Jagadeeswaran, or inferred using empirical Bayes approaches. These choices are clearly not easy to make and can have significant influence on the resulting algorithm.

The other option is of course to consider more complex priors. We agree with Stein and Hung that non-stationary priors could be a way forward, and are currently working in this direction. Another alternative to construct expressive priors would be the recent work on deep Gaussian process priors; see \cite{Dunlop2017}. Finally, we could move away from Gaussianity, allowing us more flexibility in the choice of functions on which to assign mass \citep{Cockayne2017BPNM}. Unfortunately, many of these alternatives come at the cost of additional computational requirements which may be prohibitive or require resorting to approximations which may not faithfully quantify uncertainty.

\section*{So, Should we be Bayesians for Numerical Problems?}

Regardless of the choice of prior, what is clear is that working in the space of measures is significantly harder than working in many of the function spaces commonly used in numerical analysis. As a consequence, the uncertainty quantification provided by Bayesian methods will usually come at a cost, and one should weigh this cost with the potential benefits before making a decision on whether to use a Bayesian numerical method.

A first application area where this may be useful is the field of inverse problems or computer experiments. The models studied here are often computationally expensive and based on systems of differential equations. The most popular algorithms for inference rely on repeated solution of numerical tasks such as interpolation, integration, linear algebra and optimisation. As pointed out by Stein and Hung, many of the issues facing Bayesian numerical methods, such as the choice of a prior, are also common in this literature.

Machine learning also offers many opportunities for the deployment of Bayesian numerical analysis since it is highly reliant on numerical methods and there has been a recent focus on better explaining the workings and failure modes of algorithms (that is, to provide interpretability and quantification of uncertainty). The difficulty of selecting priors can also be somewhat alleviated since we have an excellent source of prior information: the source code of the problem.

On the other hand, there is clearly no point providing uncertainty quantification for applications where uncertainty can be made negligible without excessive amounts of computation. For integration, this will usually be the case when the integrand and sampling mechanism are cheap. In this case, we would argue that being Bayesian about the numerical problem may not be necessary, or desirable.

\section*{Some Promising Research Directions}

With these application areas in mind, we conclude by highlighting areas which we believe could benefit the Bayesian approach to numerical analysis. 

The first is the ability to jointly model the solution of several quantities of interest, for which data can be obtained simultaneously or sequentially. This could for example be the solutions of several related integrals, as was considered in \cite{Xi2018MultiOutput}, but could also include other problems such as solving differential equations, linear systems, optimisation  or a mixture of these problems. In such cases, data from one of the problems could be leveraged to improve the accuracy of the algorithms approximating the other problems. Such information sharing is extremely natural within a Bayesian framework and currently underexplored in the numerical analysis literature.

Another advantage which comes out of the Bayesian viewpoint is that of chains of computations and the ability to condition on all sources of uncertainty. Take for example some of the complex Bayesian inverse problems present in the computer experiments literature. There, it is common to have to sequentially solve numerical problems in both the forward and inverse problem. Here, reducing the discretisation error in each of these problems to a negligible level may not always be possible due to computational cost. Being Bayesian about numerical problems allow us to condition jointly on the uncertainty coming from the inverse problem and the uncertainty inherent due to numerical error using Bayes' theorem. We can hence obtain posteriors which are reflect our combined uncertainty due to all sources of discretisation \cite{Cockayne2017BPNM,Oates2017hydrocyclones}. The numerical analysis of large scientific codes is not straightforward in general, but Bayesian numerical analysis provides a coherent overarching framework.

\subsubsection*{Acknowledgments}

FXB was supported by the EPSRC grants [EP/L016710/1, EP/R018413/1]. MG was supported by the EPSRC grants [EP/J016934/3, EP/K034154/1, EP/P020720/1, EP/R018413/1], an EPSRC Established Career Fellowship, the EU grant [EU/259348]. FXB, CJO, MG \& DS were all supported by The Alan Turing Institute under grant [EP/N510129/1] and/or the Lloyds Register Foundation Programme on Data-Centric Engineering.

\bibliographystyle{plainnat}
\bibliography{rejoinder_bib}

\end{document}